\documentclass[%
 aip, amsmath,amssymb,
 reprint]{revtex4-1}

\usepackage{graphicx}
\usepackage{dcolumn}
\usepackage{bm}

\usepackage[utf8]{inputenc}
\usepackage[T1]{fontenc}
\usepackage{mathptmx}

\newcommand{\ppp}{{\cal P}}
\newcommand{\sss}{{\cal S}}

\renewcommand{\Im}{\mbox{Im}\,}
\begin{document}

\preprint{AIP/123-QED}

\title{Qualitative properties of systems of 2 complex  homogeneous ODE's: a connection
to polygonal billiards}

\author{F. Leyvraz}
 \altaffiliation[Also at ]{Centro Internacional de Ciencias, Cuernavaca, Morelos M\'exico}

\email{f\_leyvraz2001@hotmail.com}

\affiliation{%
Instituto de Ciencias F\'\i sicas---Universidad Nacional Aut\'onoma de
M\'exico\\ 
Cuernavaca, 62210 Morelos, M\'exico.
}

\date{\today}

\begin{abstract}
A correspondence between the orbits of a system of 2 complex, homogeneous, polynomial 
ordinary differential equations with real coefficients and those of a polygonal billiard is displayed.
This correspondence is general, in the sense that it applies to an open set of systems of 
ordinary differential equations of the specified kind. This allows to transfer results well-known from 
the theory of polygonal billiards, such as ergodicity, the existence of periodic orbits, the absence of
exponential divergence, the existence of additional conservation laws, and the presence of discontinuities
in the dynamics, to the corresponding systems of ordinary differential equations. It also shows that 
the considerable intricacy known to exist for polygonal billiards, also attends these apparently 
simpler systems of ordinary differential equations.
\end{abstract}

\maketitle

\section{Introduction}
\label{sec:intro}
In the following, we study a special case of the system of 4 real ordinary differential equations
with homogeneous right-hand sides:
\begin{equation}
\dot{y}_i=P_i^{(r)}(y_1,y_2,y_3,y_4)\qquad(1\leq i\leq 4),
\label{general}
\end{equation}
where the $P_i^{(r)}$ are homogeneous polynomials of degree $r$ in the 4 variables $y_i$.
Such equations are a bit anomalous due to the absence of linear terms, nevertheless
they are studied to a considerable extent in various fields, such as
reaction kinetics and ecology.

The case we shall study is special in 2 respects
\begin{enumerate}
\item First, and most importantly, the system (\ref{general}) is assumed to arise from a complex 
system of 2 ODE's:
\begin{subequations}
\begin{eqnarray}
\dot{x}_1&=&p_r(x_1,x_2)\label{eq1a}\\
\dot{x}_2&=&q_r(x_1,x_2).\label{eq1b}
\end{eqnarray}
\label{eq1}
\end{subequations}
Here $p_r(x,y)$ and $q_r(x,y)$ are both homogeneous polynomials of degree $r$. 
\item Second, the coefficients of the polynomials $p_r(x_1,x_2)$ and $q_r(x_1,_2)$
are real.
\end{enumerate}
Clearly, the most important restriction is the first one. The second can certainly be
significantly generalised. However, the first one is essential, and is responsible for 
the highly atypical behaviour we identify for such systems. 

As we shall see, these systems can be completely understood, once a corresponding 
polygonal billiard problem is solved: each orbit of (\ref{eq1}) can be brought uniquely
into correspondence with the orbit of a given polygonal billiard, the characteristics
of which do not depend on the initial conditions, but only on the system (\ref{eq1})
itself. 

In particular, we find the following properties, which differ rather strongly from the 
generic properties of such ODE's: first and foremost, the dynamics is regular: that 
is, with probability one, the orbits are defined and finite for all times. As follows, for example, 
from \cite{escape}, within the sets of quadratic systems, there is an open set
with the property that the time for a solution to diverge is finite for an open
set of initial conditions. On the other hand, for systems of the type (\ref{eq1}),
solutions are non-singular for all times with probability one. Note that this 
does not mean that the orbit remains bounded: in a rather general case, as we shall see,
the orbit never diverges, but with probability one assumes arbitrarily large values
over small time intervals.

\begin{figure}[tbp]
\begin{center}
\includegraphics[scale=0.5]{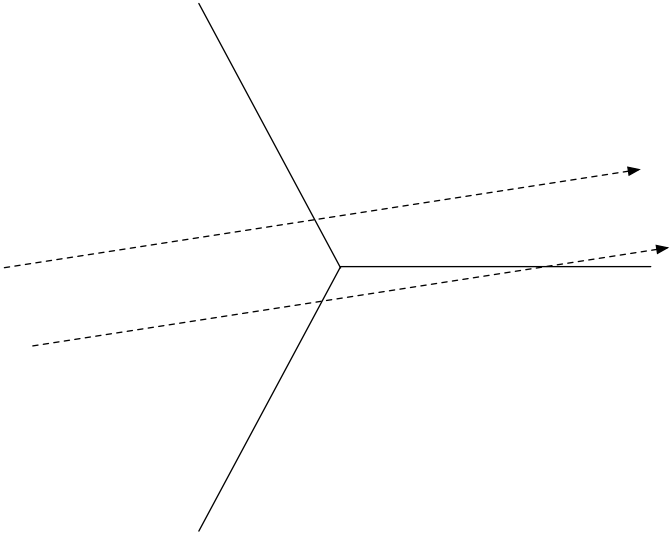}
\end{center}
\caption{ 
Two parallel orbits hitting a $2\pi/3$ corner from either side, are shown. Instead of reflecting the orbits, the
sectors are reflected. It is seen that the orbit that hits the corner on the right-hand side is reflected once,
whereas the other is reflected twice and comes off in a quite different direction. It is readily seen that this phenomenon
arises whenever the corner is not of the form $\pi/n$ for $n\in\mathbb N$.
}
\label{fig:3}
\end{figure}

In a polygonal billiard, the singular orbits (that hit a corner) have an important characteristic: 
in their vicinity, the regular orbits 
vary discontinuously: Figure \ref{fig:3} shows how this occurs for an orbit hitting a $2\pi/3$ corner, but 
the effect occurs generally, except for angles of the form $\pi/n$. In the ODE system, this is reflected in the fact
that the orbits hitting a corner diverge, so that the theorems on continuous dependence of the solution of an ODE
on initial conditions, fail in their vicinity. 

Another striking feature of these systems is the fact that their Lyapunov exponents
\cite{chaos1, chaos} are always zero. This means that the dynamics can actually 
be predicted efficiently over large times, though, as we shall see, the dynamics 
can indeed be quite complicated.

We thus show that a correspondence exists between any given system of ODE's of the type 
(\ref{eq1}) and the dynamics of a free particle bouncing elastically within a polygon, the shape of which is 
uniquely and elementarily determined by the system (\ref{eq1}). Since many results on the 
existence and nature of periodic orbits, on ergodicity, Lyapunov exponents and several other properties,
are known for polygonal billiards, it turns out that they trivially translate into corresponding properties for
the system of ODE's (\ref{eq1}). This is the essence of this paper. 

In Section \ref{sec:quad} we show how the system (\ref{eq1}) can be solved by quadratures. Since the integrals
involved are complicated, inverting them is not a trivial task, and an understanding of the orbit requires
an additional remark. In Section \ref{sec:polygon}, we establish in detail the correspondence, in Section \ref{sec:results}
we establish the general consequences of this correspondence. The results are different depending 
on whether the polygon is bounded or unbounded, so this Section is divided in 3 subsections, which treat the properties valid
in either case (Subsection \ref{subsec:genresults}) in the unbounded case, in which the orbits are 
scattering orbits (Subsection \ref{subsec:scattering}) and finally the most important csase in which the polygon is
bounded and the motion is finite(Subsection \ref{subsec:bound}). In Section \ref{sec:numerics},
we illustrate numerically some of the predictions made here, and in Section \ref{sec:conclusions}
we present conclusions ,
Finally, to keep this article self-contained,
we quote without proof the various properties of polygonal billiards used in this paper in Appendix \ref{app:a}. 

\section{Solution by quadratures}
\label{sec:quad} 
We consider the system (\ref{eq1}) of ordinary differential equations.
These are viewed as complex equations, that is, viewed as 
a system involving real quantities, they
correspond to a system of 4 equations in 4 unknowns, corresponding to the real 
and imaginary parts of $x_1$ and $x_2$.  
We rewrite these equations as follows:
\begin{subequations}
\begin{eqnarray}
\dot{x}_1&=&x_2^rP_r(x_1/x_2)\label{eq2a}\\
\dot{x}_2&=&x_2^rQ_r(x_1/x_2).\label{eq2b}
\end{eqnarray}
\label{eq2}
\end{subequations}
Here $P_r(u)$ and $Q_r(u)$ are both polynomials of degree $r$. There is a minor 
loss of generality in this description, as we assume that there exists in both equations
a term $x_2^r$. This can always be reached by a linear transformation of 
the dependent variables. By an appropriate scaling of $x_2$ we may further choose 
$Q_r(u)$ to be a {\em monic\/} polynomial. 

In the following, we present a solution by quadratures of this system. This is not new, but has been
explicitly formulated by \cite{nicklason}, and similar calculations have been performed by Garnier
\cite{garnier1, garnier2}. This approach has also been used in \cite{CCL} to identify particularly simple
special cases of (\ref{eq1}) for $r=2$.

We define the following quantities
\begin{subequations}
\begin{eqnarray}
u&=&x_1/x_2,\label{defu}\\
x_1&=&uR(u),\label{defR}\\
x_2&=&R(u).\label{defR1}
\end{eqnarray}
\label{defs}
\end{subequations}
From (\ref{eq2}) follows the following equation for $u$
\begin{equation}
\dot u=-x_2^{r-1}\left[uQ_r(u)-P_r(u)\right]=:-x_2^{r-1}S_{r+1}(u)
\label{equ}
\end{equation}
where the final equation defines $S_{r+1}(u)$, which is a monic polynomial 
of degree $r+1$. Note the use we have made of the normalisation
of $Q_r(u)$ as a monic polynomial. 

From (\ref{equ}) and (\ref{defR1}), one obtains
\begin{equation}
\dot u=-R(u)^{r-1}S_{r+1}(u).
\label{udot}
\end{equation}
Now from (\ref{eq2b}) and (\ref{defR1}) one finds
\begin{eqnarray}
\dot{x}_2&=&R^\prime(u)\dot u\nonumber\\
&=&-R(u)^{r-1}R^\prime(u)S_{r+1}(u)\nonumber\\
&=&R(u)^rQ_r(u).
\label{difficult}
\end{eqnarray}
This in turn leads to a separable equation for $R(u)$:
\begin{equation}
\frac{R^\prime(u)}{R(u)}=-\frac{Q_r(u)}{S_{r+1}(u)}.
\label{basic}
\end{equation}
Let $u_\alpha$, $0\leq\alpha\leq r$ be the zeros of $S_{r+1}(u)$, that is:
\begin{equation}
S_{r+1}(u)=\prod_{\alpha=0}^r(u-u_\alpha).
\label{Sproduct}
\end{equation}

We decompose the right-hand side of (\ref{basic})
in partial fractions, assuming that none of the $u_\alpha$ are double zeros:
\begin{equation}
\frac{Q_r(u)}{S_{r+1}(u)}=\frac1{r-1}
\sum_{\alpha=0}^r\frac{\mu_\alpha}{u-u_\alpha},
\label{partfrac}
\end{equation}
where the $1/(r-1)$ prefactor is introduced for future convenience.
Matching the $u\to\infty$ behaviours of both sides of (\ref{partfrac}), 
remembering that both $Q_r(u)$ and $S_{r+1}(u)$ are monic, we obtain
\begin{equation}
\sum_{\alpha=0}^r\mu_\alpha=r-1.
\label{sumrule}
\end{equation}
Note that the $\mu_\alpha$ and the $u_\alpha$ are altogether independent
of the initial conditions and instead characterise the system (\ref{eq1}) itself. 

(\ref{basic}) is now immediately integrated to yield
\begin{equation}
R(u)=C\prod_{\alpha=0}^r
\left(
u-u_\alpha
\right)^{-\mu_\alpha/(r-1)}
.
\label{solR}
\end{equation}
Here $C$ is an integration constant determined by the relation
\begin{equation}
x_2(0)^2=C\prod_{\alpha=0}^r
\left[
x_1(0)-u_\alpha x_2(0)\right]^{-\mu_\alpha/(r-1)}
.
\label{defC}
\end{equation}
We now proceed to a final normalisation step: the solutions of (\ref{eq1})
can always be scaled by a fixed real factor $\lambda$, which corresponds
to a scaling of $t$ by the factor $\lambda^{r-1}$. We may hence, without loss of generality, 
scale the initial conditions accordingly and therefore fix the norm of $C$. 
We thus set $|C|=1$ and
\begin{equation}
C =-e^{i{\chi_0}}.
\label{defchi}
\end{equation}
We may now determine the time-dependence of $u$ using (\ref{udot}):
\begin{equation}
\dot u= e^{i{\chi_0}} \prod_{\alpha=0}^r
\left(
u-u_\alpha\right)^{-\mu_\alpha+1}
.
\end{equation}
which leads to the expression via quadratures
\begin{equation}
 t=e^{-i{\chi_0}}\int_{u(0)}^u \prod_{\alpha=0}^r
\left(
u'-u_\alpha\right)^{\mu_\alpha-1}du^\prime
\label{solquad}
\end{equation}
where $u(0)=x_1(0)/x_2(0)$.

\section{Correspondence between the ODE's and polygonal billiards}
\label{sec:polygon}

We now limit ourselves to the subclass of systems in which the coefficients
of the polynomials $P_r(u)$ and $Q_r(u)$ are all {\em real}. 
Under these conditions, the fact that all $u_\alpha$ should be real and
simple, is no 
more exceptional. Indeed, given a system with that property, all other systems 
that are sufficiently close also have this property, so that we are in
a generic case.

The essential observation we now make is the following: if all 
$u_\alpha\in\mathbb{R}$,  all $\mu_\alpha\in\mathbb{R}$ as well. 
It then follows that, for appropriate values of the $\mu_\alpha$, specifically 
for $0\leq\mu_\alpha\leq1$, the transformation defined by (\ref{solquad})
is the conformal map from the upper half-plane to a finite, convex 
polygon $\ppp$, having $r+1$ sides and interior angles $\mu_\alpha\pi$, the well-known
Schwarz--Christoffel transformation \cite{schwarz1, schwarz2, schwarz}. It follows 
immediately from (\ref{sumrule}) that the sum of the interior angles of the polygon is, 
as it must be, equal to $(r-1)\pi$. Further note that the polygon's shape,
which is the main object of our consideration, is determined both 
by the interior angles given by the $\mu_\alpha$, and by the relative
lengths of the sides, determined by $r-3$ values of the $u_\alpha$.
We denote by $v_\alpha$ the vertices of $\ppp$ corresponding to
$u_\alpha$, by $\sss_\alpha$ the side of $\ppp$ connecting
$v_\alpha$ to $v_{\alpha+1}$, where $\alpha+1$ is computed modulo
$r+1$. To $\sss_\alpha$ corresponds in the boundary of the
upper half-plane, the interval $I_\alpha=[u_\alpha,u_{\alpha+1}]$.

Note in passing that the task we face here is different from, and in many ways 
easier than, the one usually solved by the Schwarz--Christoffel transformation: 
normally one is given a polygonal domain and looks for a conformal transformation. In that 
case, the determination of the $u_\alpha$ can be challenging \cite{schwarz1}. In our case, 
we are given the $u_\alpha$ and the angles $\mu_\alpha$, and our task is merely
to determine the image of the upper half-plane under this transformation.

For definiteness's sake, let us assume the initial condition $u(0)$ 
to be in the upper half-plane (the opposite case is similar). The map
\begin{equation}
\Phi(u)=\int_{u(0)}^udu^\prime\, \prod_{\alpha=0}^r
\left(
u'-u_\alpha\right)^{\mu_\alpha-1}.
\end{equation}
maps the upper half-plane onto the inside of a convex $r$-sided polygon $\cal{P}$
containing the origin. We have in particular
\begin{equation}
\Phi(u_\alpha)=v_\alpha\qquad(0\leq\alpha\leq r).
\label{angles}
\end{equation}

The equation (\ref{solquad}) means that the straight line $\cal{L}$ defined 
by $e^{i{\chi_0}} t$, for all real $t$, is the image of the orbit $u(t)$ under the map 
$\Phi$. 

\begin{figure}[tbp]
\begin{center}
\includegraphics[scale=0.5]{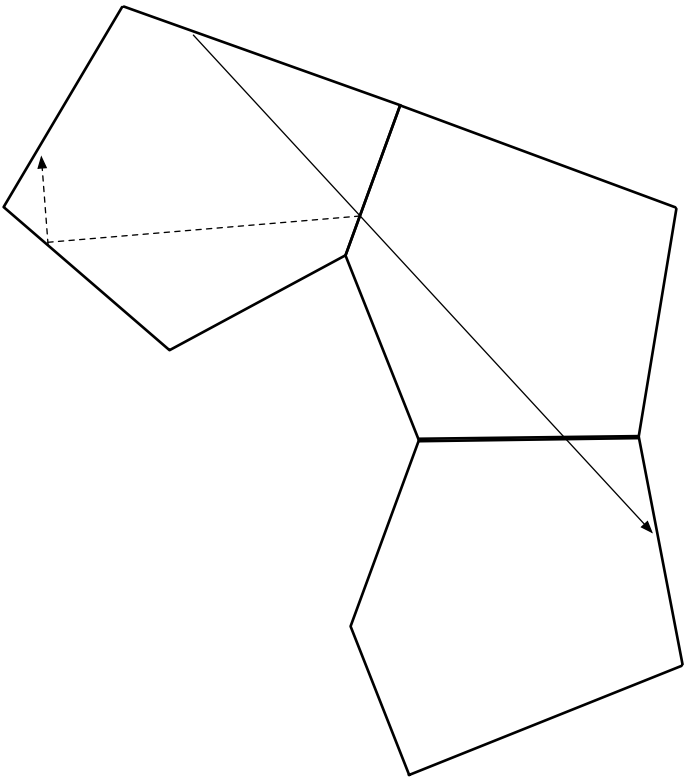}
\end{center}
\caption{ 
A straight line moves through a set of polygons, each polygon arising from the previous one via
reflection with respect to the side crossed by the line. Shown as a dashed line is the billiard orbit arising by
reflecting the straight line with respect to each side of the polygon hit by the orbit. 
}
\label{fig:ref}
\end{figure}

We must therefore determine the inverse image of $\cal{L}$ under 
$\Phi$. For the segment of $\cal{L}$ that lies entirely in $\cal{P}$, 
the corresponding part of the trajectory lies wholly in the upper half-plane.
As the line leaves $\cal{P}$ by the side $\cal{S_\alpha}$ corresponding 
to the real line interval $I_\alpha=[u_\alpha,u_{\alpha+1}]$, the corresponding orbit 
of $u$ leaves the upper half-plane by the interval $I_\alpha$. Due to the Schwarz 
reflection principle\cite{conway}, the image under $\Phi$ of the sheet which the orbit $u(t)$ enters, 
is the polygon described by the reflection of $\cal{P}$ on the side $\cal{S_\alpha}$. 
We may therefore keep the orbit of $u$ inside the upper half-plane by specularly 
reflecting it with respect to the real axis, and correspondingly keep the
line $\cal{L}$ inside the polygon $\cal{P}$, also by reflection. Indefinite repetition of 
this procedure, for both positive and negative times,
leads to a billiard orbit inside $\ppp$. For an illustration of the way this
proceeds, see Figure \ref{fig:ref}. Note that this construction
is rather similar to one used in \cite{GS} for a somewhat related problem.
If we now take the inverse image under $\Phi$ of
this billiard orbit, we obtain the orbit $u(t)$ specularly reflected each time
it crosses the real axis, from which the actual $u(t)$ orbit is readily reconstructed.

At this stage let us define some additional notation: the initial segment of the billiard trajectory
is the straight line segment $e^{i\chi_0}t$. After $n$ bounces we define the 
corresponding straight line segment to be $e^{i\chi_n}(t-\tau_n)$. Here $\tau_n$
is the time at which the orbit hits $\ppp$ and begins the $n$-th bounce.
As the billiard 
orbit is successively followed, the connection between $t$ and $u$ given by (\ref{solquad})
is modified to 
\begin{equation}
 t-\tau_n=e^{-i{\chi_n}}\int_{u(0)}^u \prod_{\alpha=0}^r
\left(
u'-u_\alpha\right)^{\mu_\alpha-1}du^\prime.
\label{solquadbill}
\end{equation}
We therefore see that the inverse image under
$\Phi$ of a billiard orbit of $\cal{P}$ 
is the orbit of $u$ reflected back into the upper half-plane each time it hits the real 
axis. Since $\Phi$ is not an easily determined map, this does not represent an exact
solution, but remembering the many results known about polygonal billiards, the 
correspondence yields several non-trivial results concerning the solutions' {\em qualitative\/} 
behaviour.

Before we proceed to describe these, however, it is of some importance to extend the 
validity of the correspondence as far as possible. In the case $0\leq\mu_\alpha<1$, the image
of the upper half-plane is an $r$-sided convex polygon. If we generalise this to 
$0\leq\mu_\alpha<2$, we obtain arbitrary bounded polygons, whether convex or not: the 
polygon's interior angles are then $\mu_\alpha\pi$, and they still add up to $(r-1)\pi$. 
The extension to negative values of $\mu_\alpha$ leads to unbounded polygons. 
Due to (\ref{sumrule}), negative values of $\mu_\alpha$ must always coexist with positive ones.
If $\mu_\alpha<0$, the integral describing $\Phi$ diverges as $u\to u_\alpha$ 
on $\mathbb{R}$, both from the right and the left. The polygon's boundary thus contains
two lines diverging to infinity and forming an angle $\mu_\alpha\pi$. We may thus draw a 
polygon corresponding to all real values of $\mu_\alpha$ satisfying both (\ref{sumrule})
and $-2<\mu_\alpha<2$. 

Further extensions, whether to values of $\mu_\alpha$ with $|\mu_\alpha|\geq2$ or complex values 
of $\mu_\alpha$ may well be possible, but it is not obvious how to extend the above construction to such cases, 
and more generally speaking, how to obtain meaningful results from them. 

\section{Consequences of the correspondence}
\label{sec:results}
\subsection{General results}
\label{subsec:genresults}
For arbitrary shapes of the polygon, the following remarks hold: generically the orbits $x_{1,2}(t)$
remain finite, since divergence could only arise if the orbit $u(t)$ hits $u_\alpha$, which does 
not happen generically. Another remarkable feature of such systems is a very sensitive
dependence on the parameters characterising the system. Indeed, the properties of polygonal 
billiards with rational and irrational angles are very different, so that the corresponding systems of
homogeneous ODE's also show such dependence. 

Another general feature is the structure of periodic orbits. In generic systems, in particular
in chaotic systems, periodic orbits are isolated. In the presence of a conservation law,
the orbits are isolated once the system is reduced to a surface where the conserved quantity 
takes a fixed value. However, for polygonal billiards, periodic orbits of even period
always appear in one-parameter families, even though no conservation law may exist,
as is the case, for instance, in irrational billiards. Again, this feature translates into 
the systems of homogeneous ODE's discussed here. 

Finally the central role played by discontinuities in the dynamics, both in polygonal billiards and in 
the homogeneous systems of ODE's we are considering here, should be emphasized. 
Whenever the orbit of a polygonal billiard hits a corner, it cannot be continued. However, as
an orbit is continuously moved through a corner, the orbits undergo a discontinuous 
variation, unless the angle of the corner is equal to an angle of the form $\pi/n$ for 
$n\in\mathbb{N}$, see Figure \ref{fig:3} for the case of a $2\pi/3$
corner. In the corresponding systems of homogeneous ODE's, hitting a corner
corresponds to divergence of the $x_{1,2}(t)$, beyond which the orbit cannot be continued,
and similarly, the orbits in the vicinity of such a divergence also show a discontinuous variation. 

\subsection{Scattering systems}
\label{subsec:scattering}
Here we consider the case in which one or more of the quantities $\mu_\alpha$ are 
negative or zero. In this case the  the polygon extends to infinity, either with straight lines 
that diverge at a strictly positive angle, or, if $\mu_\alpha=0$, two parallel sides extending 
to infinity.  The billiard orbit is then a scattering orbit in the strict sense, that is, it comes 
from infinity, bounces a finite number of times on the sides of the polygon, and then goes 
back to infinity.

The first issue we address is whether, during the scattering event, the orbit $u(t)$ may diverge. 
As is readily seen, the function $\Phi(u)$ has a well-defined finite value $\Phi_\infty$ for $u\to\infty$.
If the billiard orbit 
hits this value, $u$ will diverge for this specific value of $t$. This, of course, will generically not 
happen, but well it may occur that an orbit passes close to $t_\infty$, in which case $u$
becomes anomalously large. Indeed, for $u\to\infty$
\begin{eqnarray}
\Phi(u)&=&\int_{u(0)}^u\frac{du^\prime}{{u^\prime}^2}\, \prod_{\alpha=0}^r
\left(
1-\frac{u_\alpha}{u}\right)
^{\mu_\alpha-1}\nonumber\\
&=&
\Phi_\infty-\frac1u
\left[
1+O(u^{-1})
\right].
\label{uinf}
\end{eqnarray}
Let us now assume that $\Phi_\infty$ lies close the piece of the billiard orbit defined by $e^{i\chi_n}(t-\tau_n)$.
In other words, there exists $t_\infty\in\mathbb{C}$ such that $\Phi_\infty=e^{i\chi_n}(t_\infty-\tau_n)$
and such that $|t-t_\infty|\ll1$ on the $n$-th bounce;
$u$ therefore diverges if $t\to t_\infty$. From (\ref{uinf}) follows that, for $\Phi(u)$ close to $\Phi_\infty$, 
\begin{equation}
(t-t_\infty)u=e^{-i{\chi_n}}\left[
1+O(t-t_\infty)
\right]
\end{equation}
It thus follows that a scattering orbit that passes through $t_\infty$ has a simple pole singularity 
in $u$. Using (\ref{solR}) and (\ref{sumrule}), we see that, as $t\to t_\infty$,
\begin{eqnarray}
x_2(t)&=&R(u)\nonumber\\
&=& -e^{i{\chi_0}}\frac1u\left[
1+O(u^{-1})
\right]\nonumber\\
&=&-e^{i(\chi_0-\chi_n)}(t-t_\infty)\left[1+O(t-t_\infty)\right],
\end{eqnarray}
so that $x_2(t)$ has a simple zero, whereas $x_1(t)$ is regular at $t=t_\infty$. This divergence is
therefore not a sign of singular behaviour of the solution of (\ref{eq1}). 

\begin{figure}[tbp]
\begin{center}
\includegraphics[scale=0.5]{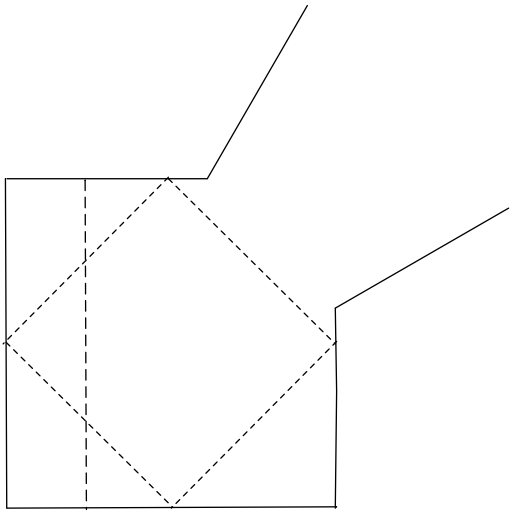}
\end{center}
\caption{ 
Example of two kinds of periodic orbits in an unbounded billiard,
which thus do not go to infinity. Note further that these 
periodic orbits are members of continuous families of periodic orbits, obtained by 
shifting the intersection with one of the sides, without modifying the initial 
velocity. These remain periodic as long as the shift is sufficiently small.
On the other hand, in the special case here shown,
almost all initial directions will eventually go to infinity
}
\label{fig:1}
\end{figure}

We now turn to the asymptotic behaviour of the scattering
orbits for large times. It is known that, for 
a large class of unbounded polygonal billiards,
{\em almost all\/} orbits eventually go to infinity, and are therefore asymptotically in
free motion. Note that this statement, while it may at first appear obvious, is
in fact quite non-trivial: see Appendix \ref{app:a} for details 
and references to the literature. The class of polygons for which it holds includes 
among others, all polygons such that $\mu_\alpha\in\mathbb{Q}$ for all $0\leq\alpha\leq r$, 
but also a set of irrational polygons large  in the sense of category\cite{category}, strictly 
speaking a denumerable intersection of dense open sets. On the other hand, the stronger
statement that {\em all\/} orbits eventually go to infinity is obviously
wrong, as shown in Figure \ref{fig:1}. Note that this example shows 
that the corresponding equations (\ref{eq1}) can have a family of periodic
orbits depending on one real parameter, as
described in the caption of Fig.~1. 

Translated into the corresponding language for the $u$ orbit, we see that, for the 
class of polygons described above, one has almost certainly
\begin{equation}
u(t)\to u_{\alpha_\pm}\qquad(t\to\pm\infty)
\label{asymptote}
\end{equation}
In the following, we limit ourselves to the 
behaviour as $t\to\infty$, but the formulae for $t\to-\infty$
are entirely similar. 

Assuming that the piece of the billiard orbit that escapes to infinity
corresponds to the $n$th bounce, we
obtain from (\ref{solquadbill}) that
\begin{widetext}
\begin{eqnarray}
t-\tau_n&=&e^{-i{\chi_n}}\int_{u(0)}^u \left[\prod_{\alpha=0}^r
\left(
u'-u_\alpha
\right)^{\mu_\alpha-1}-\left(
u'-u_{\alpha_+}
\right)^{\mu_{\alpha_+}-1}
\prod_{\alpha=0,\alpha\neq\alpha_+}^r
\left(
u_{\alpha_+}-u_\alpha
\right)^{\mu_\alpha-1}
\right]
du^\prime+\nonumber\\
&&\qquad
-\frac{e^{-i{\chi_n}}}{|\mu_{\alpha_+}|}
\left[
\prod_{\alpha=0,\alpha\neq\alpha_+}^r
\left(
u_{\alpha_+}-u_\alpha
\right)^{\mu_\alpha-1}
\right]
\left(
u-u_{\alpha_+}
\right)^{\mu_{\alpha_+}}
\label{complicated}
\end{eqnarray}
\end{widetext}
In the limit $t\to\infty$ and correspondingly $u\to u_{\alpha_+}$, the first 
summand in (\ref{complicated}) remains bounded, and is of the order
$O((u-u_{\alpha_+})^{\mu_{\alpha_+}})$, whereas the second 
diverges. Asymptotically we therefore find
\begin{subequations}
\begin{eqnarray}
t-\tau_n&=&\left[
K^{-1}(u-u_{\alpha_+})
\right]^{\mu_{\alpha_+}}\left[
1+O(u-u_{\alpha_+})
\right]\\
\label{largetimea}
K&=&-e^{i{\chi_n}} \left|
\mu_{\alpha_+}
\right|\prod_{\alpha=0,\alpha\neq\alpha_+}^r
\left(
u_{\alpha_+}-u_\alpha
\right)^{1-\mu_\alpha}
\label{largetimeb}
\end{eqnarray}
\label{largetime}
\end{subequations}
Inverting we get, since $t\to\infty$ and $\tau_n$ rermains constant
\begin{equation}
u-u_{\alpha_+}=\left(
Kt
\right)^{1/\mu_{\alpha_+}}\left[
1+O(t^{1/\mu_{\alpha_+}})
\right]
\end{equation}
This means, see (\ref{defR}), that, for large times,
$x_1(t)$ goes as $u_{\alpha_+}x_2(t)$ and that
$x_2(t)$ is given by the asymptotic expression.
\begin{eqnarray}
x_2(t)&=& e^{i{\chi_0}} \left[\prod_{\alpha=0,\alpha\neq\alpha_+}^r
\left(
u_{\alpha_+}-u_\alpha
\right)^{-\mu_\alpha/(r-1)}\right]
\left(
Kt
\right)^{-1/(r-1)}\times\nonumber\\
&&\qquad\left[
1+O(t^{1/\mu_{\alpha_+}})
\right]
.
\label{largetimes}
\end{eqnarray}
The leading behaviour is readily understood in elementary terms: if $u\to u_{\alpha_+}$
as $t\to\infty$, in this limit, (\ref{eq2b}) yields
\begin{equation}
\dot{x}_2\simeq x_2^rQ_r(u_{\alpha_+}).
\end{equation}
The leading term thus involves only the rudimentary power-law behaviour following from 
the equations' homogeneous nature. However, the subleading term depends non-trivially on 
the geometry of $\ppp$. In particular, it depends on the angle at which the infinite channel
diverges, so that these subleading exponents will in general differ for $t\to\infty$ and $t\to-\infty$,
if namely the orbit enters through one channel and leaves by another, which well may happen. 

We may also sketch what happens in the limiting case in which the polygon $\cal P$ is unbounded 
because two of its adjacent sides are parallel. In that case, again the orbit almost surely goes 
to infinity for the same class of billiards as stated above. In that case, however, the orbit generically 
goes to infinity bouncing between the two parallel sides infinitely often. Let the two 
parallel sides of $\ppp$ be $\sss_{\alpha-1}$ and $\sss_\alpha$, the vertex at infinity 
being $v_\alpha$. The infinite set of bounces on 
$\sss_{\alpha-1}$ and $\sss_\alpha$ correspond to the $u$ orbit crossing the intervals
$I_{\alpha-1}$ and $I_\alpha$ alternately infinitely often, and the approach of $t\to\infty$
corresponds to $u\to u_\alpha$. An elementary extension of the above calculations 
shows that this approach is exponentially rapid. 

Finally, let us point out that we have limited ourselves here to a rather special kind of unbounded polygon,
namely a finite polygon connected to infinity by channels. Other cases are quite possible: one can, for instance
readily generate the {\em outside\/} of a finite polygon as the image of the upper half-plane
via an appropriate Schwarz--Christoffel transformation. Such cases are, however, even simpler: no periodic
orbits arise, and all orbits go from infinity to infinity after a finite number of bounces. 

\subsection{Bound systems}
\label{subsec:bound}

The main statement for bound systems is that $u(t)$ remains bounded apart from the
possible divergences linked to passing through the point $\Phi_\infty$ as well 
as when the billiard orbit hits a corner of the polygon $\cal P$. However, as we have noted
before, the former divergence does not correspond to a singularity of the dynamics
(\ref{eq1}), so we may say that $x_1$ and $x_2$ remain bounded unless the orbit
hits, or comes close to, a corner. If it hits a corner, we cannot proceed further.
On the other hand, in the vicinity of a corner, we must determine the behaviour of 
the solution of (\ref{eq1}).

Being in the vicinity of a corner is equivalent, for the orbit of $u$, to being 
in the vicinity of an $u_\alpha$, say $u_{\alpha_0}$. It follows that, if the orbit
in the $n$-th bounce hits the corner $v_{\alpha_0}$ corresponding to 
$u_{\alpha_0}$ at a time $t_{\alpha_0}\in\mathbb{C}$, that is, if
\begin{equation}
e^{i\chi_n}(t_{\alpha_0}-\tau_n)=\Phi(u_{\alpha_0})=v_{\alpha_0},
\label{cornerposition}
\end{equation}
we may again derive (\ref{complicated}) with $\alpha_+$ replaced by $\alpha_0$. As above, one obtains
\begin{subequations}
\begin{eqnarray}
t-t_{\alpha_0}&=&\left[
K^{-1}(u-u_{\alpha_0})
\right]^{\mu_{\alpha_0}}\left[
1+O(u-u_{\alpha_0})
\right]\\
\label{cornersinga}
K&=&-e^{i{\chi_n}} \left|
\mu_{\alpha_0}
\right|\prod_{\alpha=0,\alpha\neq\alpha_0}^r
\left(
u_{\alpha_0}-u_\alpha
\right)^{1-\mu_\alpha}
\label{cornersingb}
\end{eqnarray}
\label{cornersing}
\end{subequations}
Inverting, it follows that
\begin{equation}
u-u_{\alpha_0}=
\left(
K
\left|
t-t_{\alpha_0}
\right|
\right)
^{1/\mu_{\alpha_0}}
\left[
1+O
\left(
\left|
t-t_{\alpha_0}
\right|^{1/\mu_{\alpha_0}}
\right)
\right].
\label{corner1}
\end{equation}
The connection between $t$ and $u$ near a corner
is thus given by the power $1/\mu_{\alpha_0}$. 
For $x_2(t)$ one finds
\begin{eqnarray}
x_2(t)&=&-e^{i\chi_0}\left(
u-u_{\alpha_0}
\right)^{-\mu_{\alpha_0}/(r-1)}\prod_{\alpha=0;\alpha\neq\alpha_0}^r\left(
u-u_\alpha
\right)^{-\mu_\alpha/(r-1)}\nonumber\\
&=&K^\prime 
\left|
t-t_{\alpha_0}
\right|^{-1/(r-1)}\left[
1+O
\left(
\left|
t-t_{\alpha_0}
\right|^{1/\mu_{\alpha_0}}
\right)
\right].
\label{x2corner}
\end{eqnarray}
Here $K^\prime$ is another constant. 
Since $u(t)$ is close to the finite value $u_{\alpha_0}$, it follows that in the vicinity
of the corner, $x_1(t)$ behaves as $u_{\alpha_0}x_2(t)$, so that the 2 variables have
similar qualitative behaviour, unless, of course $u_{\alpha_0}=0$, in which case the leading behaviour 
of $x_1(t)$ is the subleading behaviour of $x_2(t)$. 

The behaviour upon hitting a corner corresponds to the case in which $t_{\alpha_0}\in\mathbb{R}$, so that
the values of $x_{1,2}(t)$ diverge. On the other hand, coming close to a corner means that 
$\Im t_{\alpha_0}$ is small. In this case, the values of $x_{1,2}(t)$ become large, and their subsequent behaviour depends
on the sign of the imaginary part, a fact which corresponds to the discontinuity of the billiard
dynamics near a corner.

We may now apply well-known results for polygonal billiards to obtain corresponding
results for the dynamics of (\ref{eq1}). 

For arbitrary polygons, we cannot say very much. The most important generally valid
result is the existence of singular orbits and the fact that they have measure zero. Indeed, whenever
an orbit in $\ppp$ hits a corner $v_\alpha$ such that $\mu_\alpha\neq1/n$, with $n\in\mathbb{N}$, 
the orbit cannot be continued continuously: this arises from the fact that an orbit which comes
arbitrarily close to $v_\alpha$ but hits first the side $\sss_{\alpha-1}$ and then $\sss_\alpha$,
comes out at a different direction from a similar orbit which first hits  $\sss_\alpha$  and then 
 $\sss_{\alpha-1}$. {\em Only\/} in the specific case $\mu_\alpha=1/n$, for $n\in\mathbb{N}$, does 
continuity hold; the dynamics can then be meaningfully defined after hitting a corner. However, there
are only a finite number of cases in which {\em all\/} $\mu_\alpha$ satisfy this condition, and
these correspond to integrable cases \cite{garnier1}. 
However, the dynamics remains well-defined, since the orbits hitting a corner form a set of zero 
measure. 

The discontinuities are nevertheless important, and may be said to structure the entire 
set of orbits. In the case of ergodic billiards, see below, almost all orbits pass arbitrarily close
to a corner, and two arbitrarily close orbits may be eventually separated 
by hitting a corner on different sides. 

If, on the other hand, two orbits differ initially by an infinitesimal amount, the rate of divergence 
of the distance between the two orbits is linear, that is, the Lyapunov exponents \cite{chaos1, chaos} are all zero. 

Another universally valid remark is the following: whereas it is not known whether any given 
irrational polygon has a periodic orbit, it is known that {\em when\/} it does, 
the orbits with a primitive period consisting of an even number of bounces 
all appear in one-parameter parallel families: that is, if the orbit is shifted by 
a sufficiently small amount without changing its direction of motion, the orbit remains periodic. 
As the orbit is shifted further, it will typically disappear by hitting a corner.
This is, of course, in in clear contrast to the behaviour of generic or chaotic systems. 

It should be added that proving results for arbitrary polygons is quite difficult. Numerical work 
thus provides important additional indications. For valuable results obtained in this
manner, see in particular \cite{numtrian}.

On the other hand, if the $\mu_\alpha$ are rational, we can additionally
say the following: 
\begin{enumerate}
\item The angle at which the orbit hits the boundary
can only take finitely many different values. Since the map $\Phi$
is conformal, we see that, whenever $u(t)$ crosses
the real axis, $\arg\dot u(t)$ can only take finitely many values.
\item The polygonal billiard has a dense set of periodic
orbits: this again translates into the corresponding statement for the $u(t)$ orbit. 
\item With probability one, an initial direction is ergodic, in the sense that the 
points where the orbit hits $\ppp$ are {\em uniformly distributed\/} on $\ppp$. This implies 
that the intersections of the orbit $u(t)$ with the real axis are uniformly distributed with 
respect to the density
obtained fro the uniform measure by taking the inverse image of  $\ppp$ under $\Phi$. 
\item An interesting remark also follows from the theorem that any orbit which hits one of the sides
of a rational polygon at a right angle, is periodic, see Appendix \ref{app:a} for details.  Since the 
Schwarz--Christoffel map is conformal, this translates into the statement that any solution which 
hits the real axis perpendicularly is periodic. This means in particular that whenever
the $\mu_\alpha\in\mathbb{Q}$ lead to a bounded $\ppp$, the initial condition
$\chi_0=\pi/2$ leads to a periodic orbit for all values of $u(0)$ for which the orbit is non-singular, and
hence for almost all values of $u(0)$.
\end{enumerate}
\begin{figure}
\begin{center}
\includegraphics[scale=0.4]{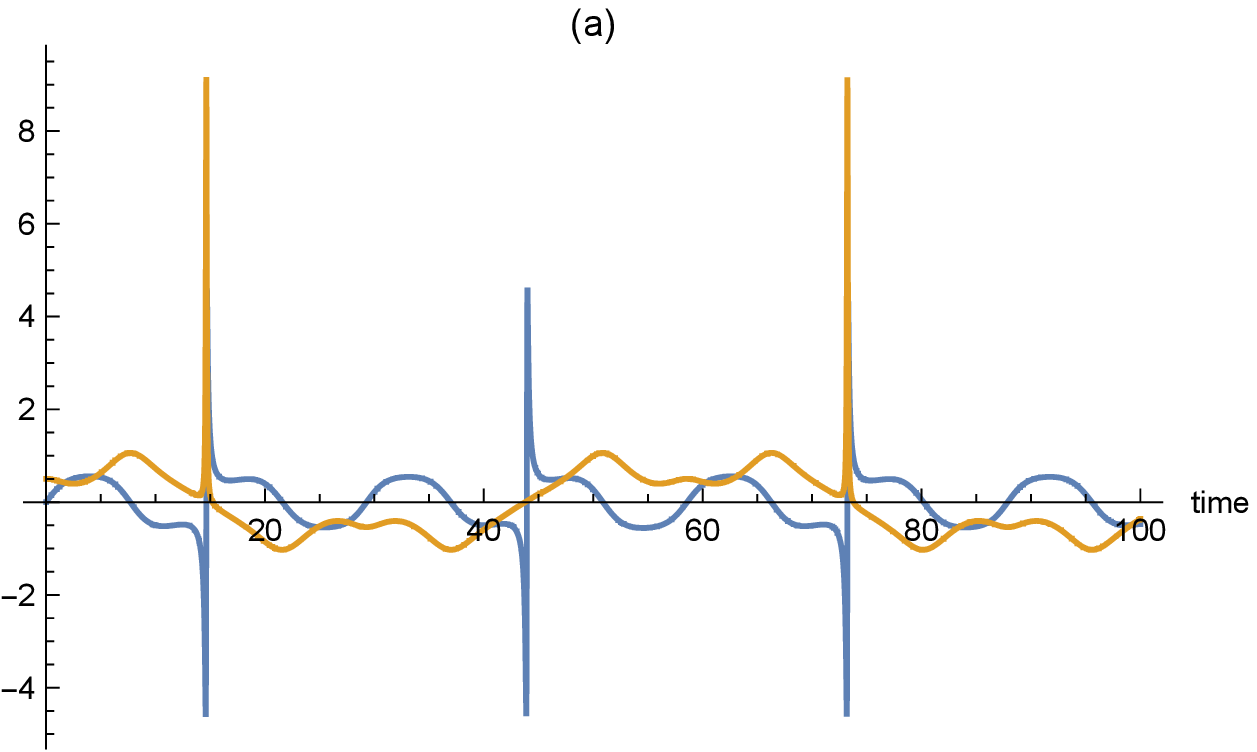}\includegraphics[scale=0.4]{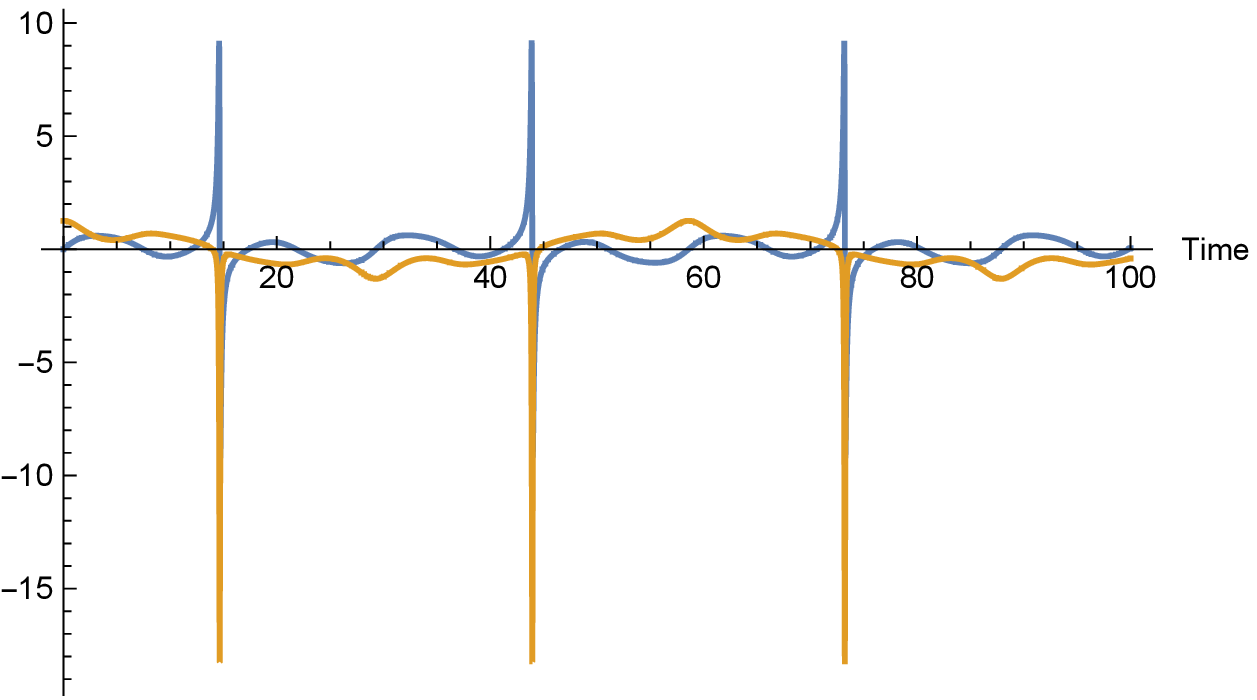}
\includegraphics[scale=0.5]{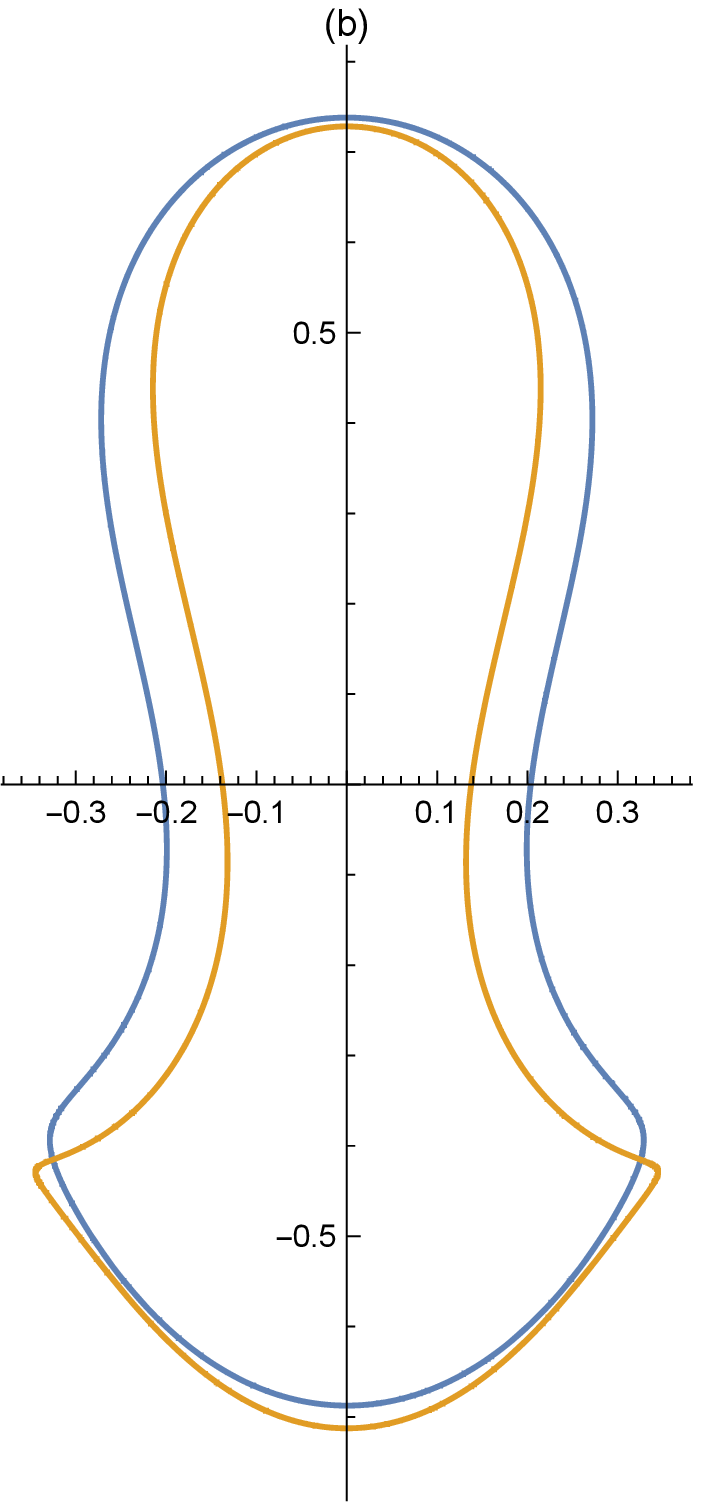}
\includegraphics[scale=0.5]{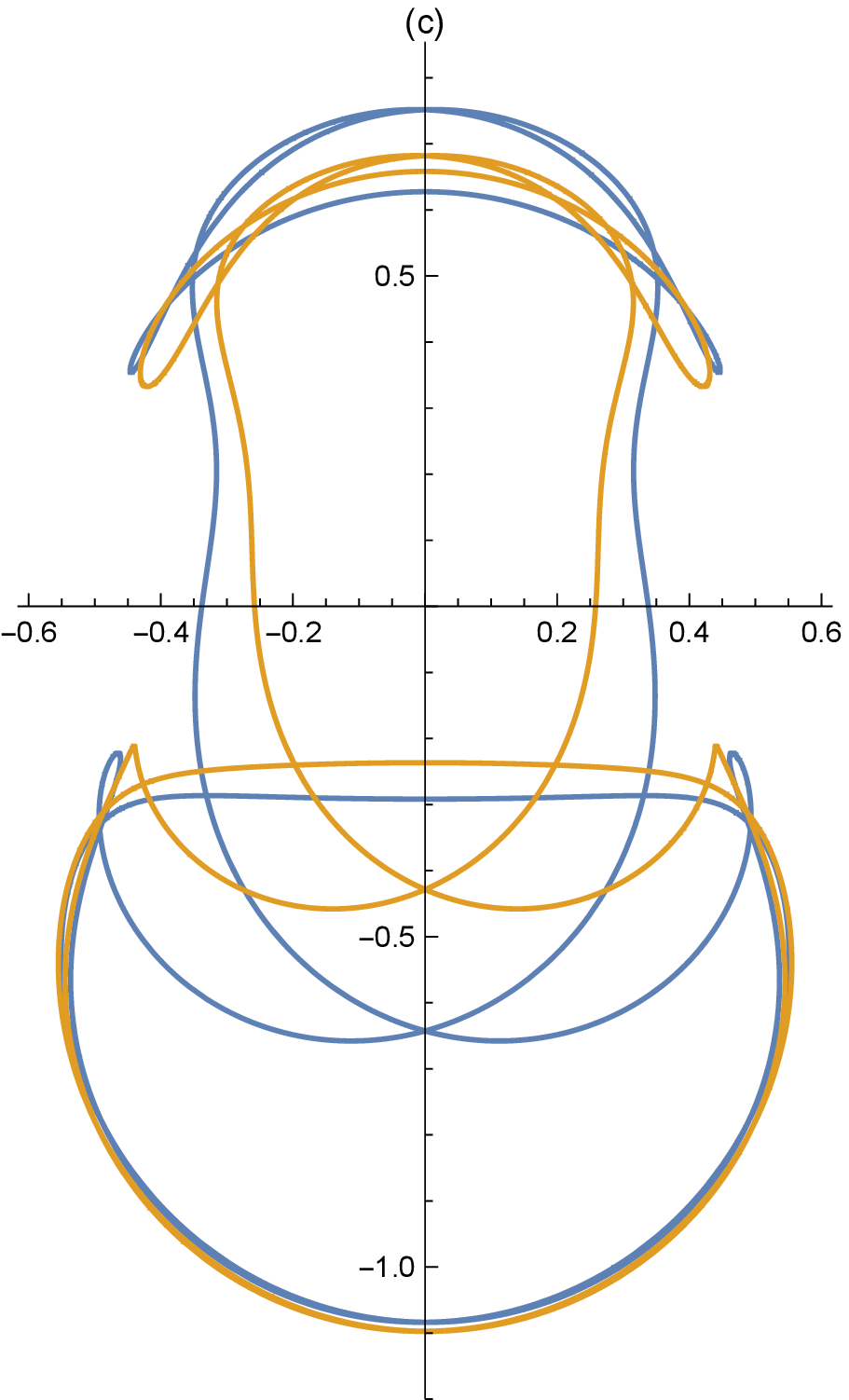}
\end{center}
\caption{ 
Various periodic orbits obtained by taking $\chi_0=\pi/2$ and $u('0)$ real. (a) shows the real and imaginary parts
of both $x_1(t)$ and $x_2(t)$, displaying the solution's complete periodicity, for $u(0)=0.4$. Parts (b) and (c)
show the orbit of $x_1(t)$ in the complex plane for the values of $u(0)$ equal to $0.46$ and $0.47$ in (b)
and $0.53$ and $0.55$ for (c). Bth the similarity and the difference in the different cases are clear. 
}
\label{fig:5}
\end{figure}

Finally let us discuss the issue of the orbit's boundedness. When the billiard is bounded, then 
clearly so is the billiard orbit. The inverse image of the triangle via the map $\Phi$
yields $u$ and $R(u)$ yields $x_2(t)$. The only way in which $x_2(t)$ can diverge,
is if $u$ takes the values $u_\alpha$, $0\leq\alpha\leq r$, which themselves correspond
to the corners of the triangle. Whenever the triangle is ergodic, that is, if the triangle is
rational, or if it belongs to the large set of ergodic irrational triangles, then almost every orbit 
passes arbitrarily close to a corner. More specifically, for almost every orbit we may state 
that the average fraction of time spent within a distance $\epsilon$ of a corner 
is itself proportional to the area of the $\epsilon$ neighbourhood of
the corner, that is $\epsilon^2$. On the other hand, passing within a distance 
$\epsilon$ of a corner means that $x_{1,2}(t)$ are of order $\epsilon^{-1/(r-1)}$. 
Thus, for any $B$ sufficiently large, the average fraction of time such that $|x_{1,2}(t)|>B$
goes as $B^{-2(r-1)}$. Qualitatively, this means that sudden sharp peaks of the solution 
will occur rather frequently, and that the probability of $x_{1,2}(t)$ taking large values 
decays as a power-law. The appearance of sharp peaks is indeed frequently observed in 
numerical work, see for example the periodic orbit in Figure \ref{fig:5}, which were 
not selected for the purpose.

\section{Numerical illustrations}
\label{sec:numerics}

In the following we illustrate using numerical simulations some of the findings described in
Section \ref{sec:results}. All the simulations are performed directly on the system (\ref{eq1}),
without using the results of Section \ref{sec:quad}. 

The system is constructed from the given data $\mu_\alpha$, $1\leq\alpha\leq r+1$,
which vary from system to system,  as follows:
the $u_\alpha$ are always conventionally taken to be
\begin{equation}
u_\alpha=\alpha-1/2-\left\lfloor{\frac r 2}\right\rfloor
\end{equation}
and the polynomial $S_{r+1}(u)$ is computed from (\ref{Sproduct}), from which $Q_r(u)$
and from that eventually $P_r(u)$ are computed using (\ref{partfrac}). 
The initial conditions are taken  with a random, or generic, value of $\chi_0$ and, 
if not otherwise stated, with a value 
of $u(0)=1/4$ always different rom $u_\alpha$. Since $|C|=1$, we can fully determine the initial conditions. 
If not stated otherwise, we shall always be dealing with the case $r=2$.

First let us show periodic orbits. As we saw, whenever the $u_\alpha$ are real and $\chi_0=\pi/2$,
the resulting orbit is almost surely periodic. Further, they vary continuously as $u(0)$ varies, apart from an 
obvious discontinuity when $u(0)$ crosses a $u_\alpha$, since this corresponds to a corner.
We show this in Figure \ref{fig:5}, where we look at a a rational case  $\mu_0=1/2, \mu_1=3/7$ and $\mu_2 =3/14$
which yields the equations
\begin{eqnarray}
\dot{x}_1&=&\frac3{14}x_1^2+\frac5{14}x_1x_2-\frac38x_2^2,\\
\dot{x}_2&=&x_1^2-\frac97x_1x_2+\frac3{28}x_2^2.
\label{rationalex}
\end{eqnarray}

\begin{figure}
\begin{center}
\includegraphics[scale=0.7]{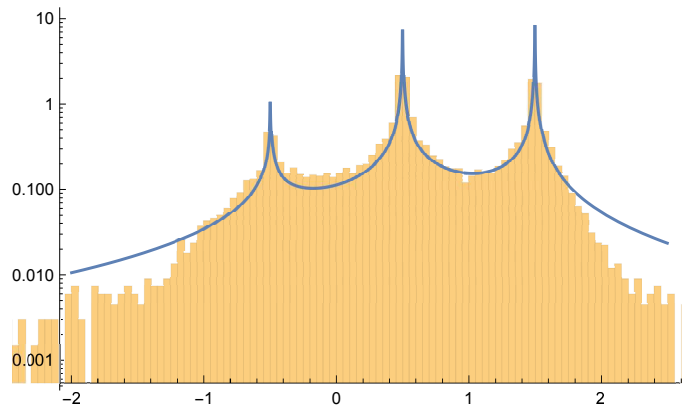}
\end{center}
\caption{ 
Histogram of the positions of the crossings of the orbit of the real axis, together with the 
prediction assuming ergodicity of the corresponding billiard orbit. This corresponds to 
the case $\mu_0=1/2, \mu_1=3/7$ and $\mu_2 =3/14$, and
$\chi_0=2.51558$. Shown is one single orbit of length $5\cdot10^4$. Note that these 
integrations were performed with Mathematica 11 with a precision of 50 decimals. Using the
standard precision leads to strong deviations from the predictions, and even 30 decimals 
are not quite satisfactory. This is possibly due to systematic errors in the treatment
of the divergence near the discontinuity caused by a corner. The systematic underestimate for
large values of $|u|$ may possibly still be such an effect. 
}
\label{fig6}
\end{figure}

We proceed to display ergodicity. If the triangle's angles, that is the $\mu_\alpha$, are rational, 
then almost every direction is ergodic. The places where the orbit is reflected on the triangle
 $\ppp$ are thus uniformly distributed. Translating this to the $u$ variables,  this means that 
 the values of $u$ where the orbit crosses the real axis have the probability distribution
\begin{equation}
p(u)=\frac1{\cal N}\prod_{k=0}^r(u-u_\alpha)^{\mu_\alpha-1},
\label{prodPhi}
\end{equation}
where $\cal N$ is the normalisation.
An example of the histogram for the $u$ values of the real crossings of a single orbit over a time of
$5\cdot10^4$ is given, together with the predicted distribution (\ref{prodPhi}). We see a good agreement in the case 
described by Figure \ref{fig6} in the rational case shown in (\ref{rationalex}). A similarly good 
agreement (not reported) is found for the irrational case $\mu_0=1/\sqrt5$, $\mu_0=1/\sqrt7$ and 
$\mu_2=1-1/\sqrt5-1/\sqrt7$.

\begin{figure}
\begin{center}
\includegraphics[scale=0.7]{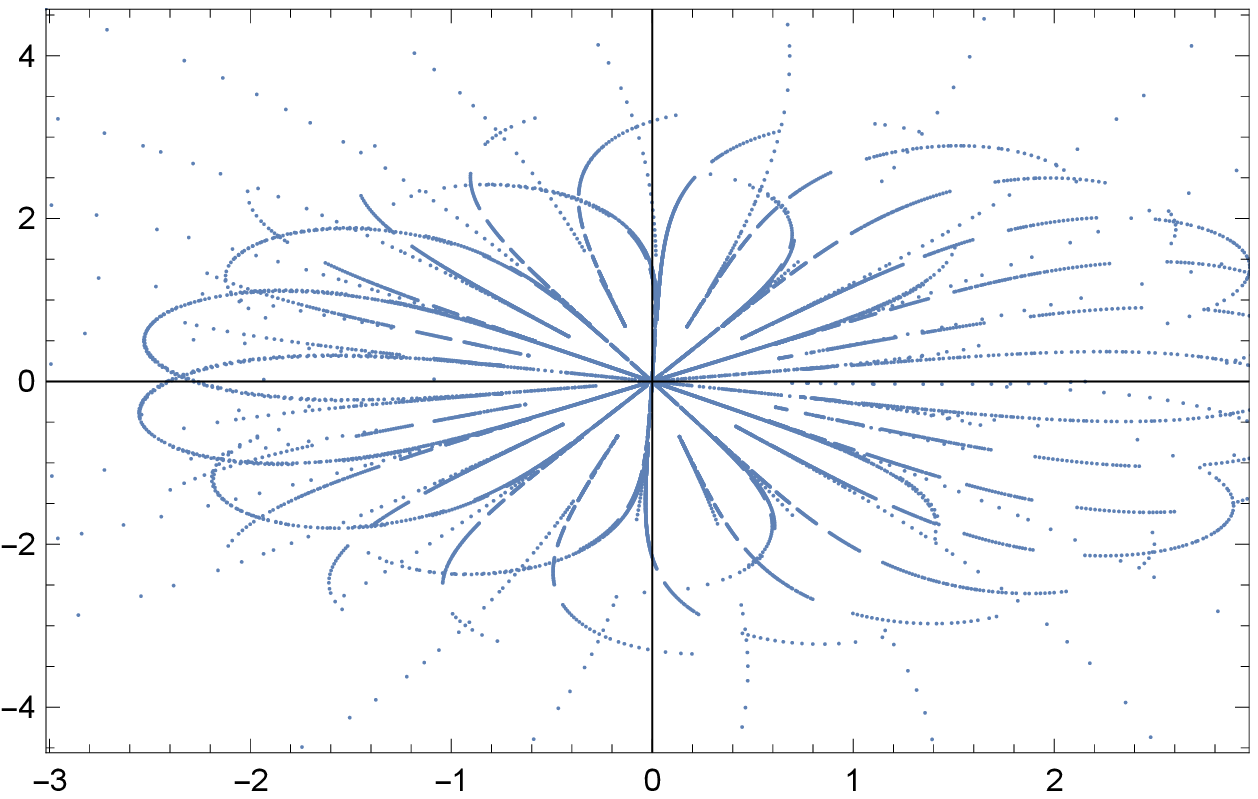}
\includegraphics[scale=0.7]{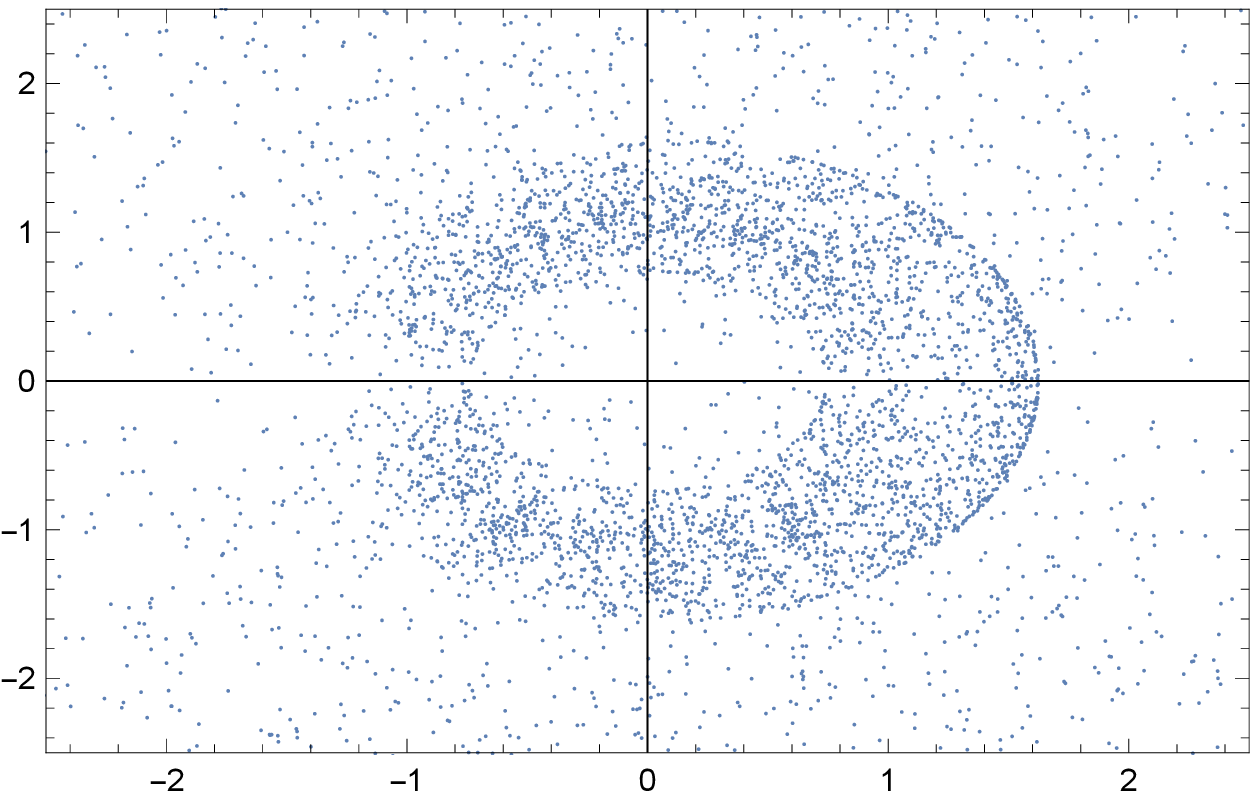}
\end{center}
\caption{ 
Poincar\'e plots for rational (above) and irrational (below) values of $\mu_\alpha$.
Specifically, the points represent the complex values of $x_2(t)$ at the times when
$u(t)=x_1(t)/x_2(t)$ crosses the real axis, where one single orbit of duration $5\cdot10^4$
starting with $u(0)=1/4$ and $\chi_0=2.51558$. 
}
\label{fig7}
\end{figure}

On the other hand, in Figure \ref{fig7}, we display evidence for one of the basic differences between 
rational and irrational angles: in Figure \ref{fig7},
we display the values of $x_2(t)$ on the complex plane at those times in which 
$u(t)=x_1(t)/x_2(t)$ crosses
the real axis for one single long orbit. Indeed, an arbitrary orbit lies in the 3-dimensional 
subspace of possible values of $x_1$ and
$x_2$ defined by the equation $|C|=1$. The intersection of the orbit with the 2-dimensional space
defined by imposing the additional condition $\Im x_1(t)/x_2(t)=0$ are thus isolated points. 
Apart from this we have no further indication. In the two plots of this nature shown in Figure \ref{fig7}, 
the lower one, corresponding to the case of irrational values of $\mu_\alpha$, indeed shows a set of 
points more or less randomly scattered on the plane. However, this is definitely not the case for the 
upper diagram, which corresponds to simple rational values of the $\mu_\alpha$. There the set of 
points is essentially a set of curves, in other words, it is one-dimensional. This corresponds, of
course, to the fact that in the corresponding orbit in the triangular billiard, the direction of the orbit 
can only take a finite number of values \cite{trian1}.

\begin{figure}
\begin{center}
\includegraphics[scale=0.7]{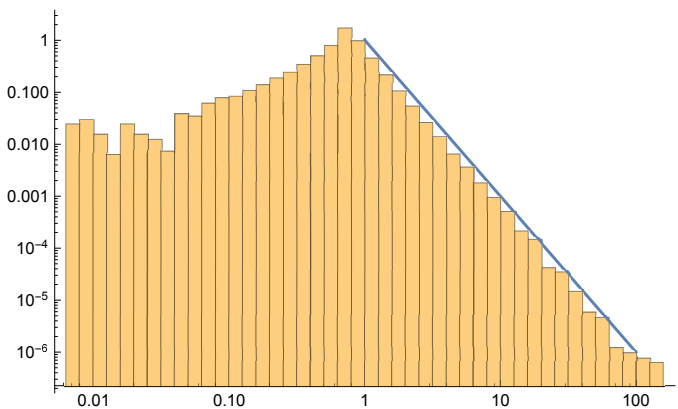}
\end{center}
\caption{ 
Histogram of the absolute values of $x_1(t)$ taken over an orbit of 
duration $5\cdot10^4$. Note a clear power-law decay, due to the large values 
of $x_{1,2}(t)$ arising when the corresponding billiard orbit approaches a corner. 
The continuous curve corresponds to the $B^{-3}$ decay predicted by theory. 
}
\label{fig8}
\end{figure}

Finally, let us verify the validity of the remarks made at the end of Section \ref{sec:results} concerning
large peaks in the values of $x_{1,2}(t)$. We consider the case $r=2$, and the rational case discussed above.
We find as a histogram of the absolute value of $x_1(t)$ taken at unit time intervals for an orbit
of duration $5\cdot10^4$. The existence of a power-law is undeniable, and the agreement 
with the theoretical prediction of an exponent $-3$ is fairly convincing.

\section{Conclusions}
\label{sec:conclusions}

We identify a class of systems of 2 complex ODE's  with remarkable properties which 
follow from the fact that we can associate to every orbit of the system a unique orbit
of a corresponding bounded polygonal billiard. From this identification follow various remarkable 
qualitative properties: the Lyapunov exponents are all zero, the motion almost surely 
never diverges and remains bounded by a constant $B$ for a fraction of the time that 
goes to one as $B\to\infty$.

While the results are rather special, being limited to systems of 4 real 
homogeneous ODE's derived from 
a complex analytic system, they can be significantly extended: since the 
properties here described are qualitative in nature, they extend to every system that 
can be obtained from (\ref{eq1}) via a change of variables. As a trivial example, using 
real linear transformations, it is possible to obtain homogeneous systems of 4 ODE's for
which the analyticity property is hidden. Similarly, all non-linear transformations
which preserve the homogeneity property can also be used to extend the relevant
class. 

Similarly, starting from the complex Newtonian equation
\begin{equation}
\ddot z=z^k
\end{equation}
we obtain by the transformation\cite{CCL}
\begin{eqnarray}
x_1&=&z^{(k-1)/2},\\
x_2&=&\frac{\dot x_1}{x_1}.
\end{eqnarray}
the set of complex equations
\begin{eqnarray}
\dot{x}_1&=&x_1x_2,\\
\dot{x}_2&=&x_1^2+\frac2{1-k}x_2^2.
\end{eqnarray}
These belong to our class for all real values of $k\neq1$, so the various results derived above, 
concerning the existence of periodic orbits, the vanishing of the Lyapunov exponent
and so on, all follow for this Newtonian equation. Note that, in this case, the existence 
of a solution in terms of quadratures follows trivially from energy conservation, but this
solution leads to hyperelliptic integrals for which it is not straightforward to obtain the various
results stated above.

Other extensions are possible. In particular, it is possible to extend the solution by quadratures
to the case of a set of 2 homogeneous complex ODE's with a linear term of the following form
\begin{subequations}
\begin{eqnarray}
\dot{x}_1&=&-\alpha x_1+p_r(x_1,x_2)\label{eq1lineara}\\
\dot{x}_2&=&-\alpha x_2+q_r(x_1,x_2).\label{eq1linearb}
\end{eqnarray}
\label{eq1linear}
\end{subequations}
However, the nature of the billiard motion is significantly different and 
its study is left for future work.

\section*{Acknowledgements}

I would like to acknowledge financial support by 
UNAM PAPIIT-DGAPA-IN113620 as well as by CONACyT Ciencias B\'asicas 
254515.
\section*{AIP Publishing Data Sharing Policy}

Data sharing not applicable---no new data generated.

\appendix

\section{Known results on triangular billiards}
\label{app:a}

Here we summarise the results known on polygonal billiards which we use in this paper. 
To avoid unnecessary complications, we define a polygonal billiard to be a particle
moving with unit velocity inside a bounded polygon, and being specularly reflected whenever the
trajectory hits a side of the polygon. We  limit ourselves to simply connected polygons (no ``holes'')
as these are the only ones generated by the Schwarz--Christoffel transformation as we use it. 

As an aside, note that polygonal billiards are different in one important respect from ordinary 
dynamical system: there exist orbits for which no continuation is possible past
a given point, namely when they hit one of the polygon's vertices. Additionally, two orbits 
that are initially close to each other, but hit a vertex on different sides, are in general
separated by a finite amount afterwards: in other words, the dynamics is discontinuous. However, 
the set of singular orbits, namely those which encounter one or two vertices in their course, is denumerable,
and hence of measure zero. 

We divide this in
two essentially different parts: the results concerning rational billiards, that is, billiards
such that all their interior angles are rational multiples of $\pi$ and those concerning
arbitrary billiards, which will generally be assumed not to be rational. Note that for an {\em unbounded\/}
rational polygonal billiard, we include a requirement that all the angles at infinity be rational. 

The basic property distinguishing rational billiards from others is the existence of an 
additional conservation law: any orbit on a rational billiard can only assume a finite number 
of different velocities, or said differently, it can only go in a finite number of different directions. 
The main results are the following:
\begin{enumerate}
\item Periodic orbits always exist, and the set of directions corresponding to periodic orbits is dense\cite{trian4, trian1}.
\item The set of directions for which the intersections with the polygon $\ppp$ are not dense 
in $\ppp$ is denumerable\cite{trian1}. Note, however, that to each direction there may correspond an interval
of parallel orbits.
\item The set of directions $\theta$ for which the directional dynamics is not ergodic, has 
measure zero\cite{trian3, trian1}. By a direction $\theta$ being ergodic we mean the following: 
let $f(s)$ be an arbitrary continuous function of the arclength of $\ppp$, and the total length of $\ppp$ 
be normalised to 1. If $x_n$, $1\leq n<\infty$ are the successive points at which an arbitrary 
orbit having direction $\theta$, intersects $\ppp$, then ergodicity implies
\begin{equation}
\lim_{N\to\infty}\frac1N\sum_{k=1}^Nf(x_k)=\int_\ppp f(s)ds.
\end{equation}
\item Any orbit that hits any side at a right angle is periodic. Indeed, it can be shown that, 
an orbit with such an initial
condition will eventually return to the same side, again hitting it perpendicularly. This then 
automatically leads to the orbit backtracking on itself, thus becoming eventually periodic. 
Note that this result may well hold under less restrictive conditions: it holds, for example, 
for arbitrary right triangles, whether rational or not, as well as for all polygonal billiards, the sides of which
are all parallel to one or the other of two directions and has numerically been found to hold 
generally\cite{numtrian}. 

\end{enumerate}

\begin{figure}[tbp]
\begin{center}
\includegraphics[scale=0.5]{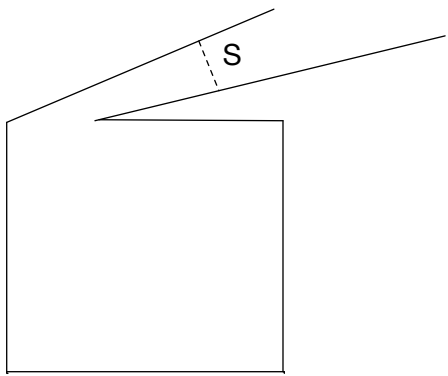}
\end{center}
\caption{ 
An unbounded billiard with one channel sealed off from a finite part: once the particle
crosses the dotted line $S$, it will never come back to the finite part. An orbit that remains forever
in the finite part, can thus never hit $S$.
}
\label{fig:2}
\end{figure}

Note finally that the above results easily imply the claim that, for all unbounded rational 
billiards, almost all orbits will go to infinity: indeed, we may seal off all the unbounded
channels by adding one wall that separates each channel from an inside finite region.
The additional separating wall can additionally be put in such a way that, whenever a particle
crosses the wall, it necessarily gets into the channel with no possibility of returning
to the finite region, see Fig.~\ref{fig:2} for an illustration. An orbit that remains forever 
in the finite region can never hit these sides of the finite region that separate it from 
the infinite channels. By Property 3 the set of corresponding directions is denumerable, 
and thus the set of such orbits has measure zero. 

For more general polygons, the results are very different. In particular, it is no more true that 
each orbit only goes in a finite number of directions. The main result is then that the set of 
ergodic billiards, where ergodic is now taken, as usual, to refer both to velocity and position,
is a large set, in the sense of being the countable intersection of dense open subsets of the 
space of all $n$-sided polygons. To define the latter, we assume that the set of such polygons
is normalised so that all polygons have perimeter one. The polygons are then determined 
by a finite number of parameters (angles and sides) all of which remain bounded. The set of
all $n$-sided polygons is thus an open bounded set in a finite dimensional space, so that 
topological concepts can be defined. It is not known at present whether this set has positive 
measure. 
 
 A general, rather obvious
property of polygonal billiards, is that their Lyapunov exponent is zero\cite{trian1}. 
Concerning the existence of periodic orbits, rather little is known. Whereas it is assumed 
that all triangles have periodic orbits, this is only known with certainty for triangles, the 
largest angle of which is less than or equal to 100 degrees \cite{schwartz1,schwartz2}. 
Additionally, it is shown in \cite{schwartz1} that the minimal number of bounces 
for a periodic orbit is not continuous as a function of the angles of the triangle: indeed, 
it diverges in the vicinity of the right triangle with angles $(\pi/2,\pi/3, \pi/6)$. The problem 
is therefore unexpectedly difficult.

\bibliographystyle{unsrt}

\begin{thebibliography}{99}

\bibitem{escape} W. M. Getz and D. H. Jacobson, Sufficiency Conditions 
for Finite Escape Times in Systems of Quadratic Differential Equations, 
J. Inst. Maths. Applics. (1977) {\bf19}  377--383 (1977)

\bibitem{chaos1} G. Benettin, L. Galgani, A. Giorgilli, and J. M. Strelcyn, 
Tous les nombres caract\'eristiques de Lyapounov sont effectivement
calculables, C.R. Acad. Sc. Paris, {\bf286A} 431--433
(1978).

\bibitem{chaos} G. Benettin, L. Galgani, A. Giorgilli, and J. M. Strelcyn, 
Lyapunov characteristic exponents for smooth dynamical systems: a 
method for computing all of them, Meccanica {\bf15} (1) 9--20 (1980). 

\bibitem{nicklason} G. R. Nicklason, The general phase plane solution of the 2d
homogeneous system with equal Malthusian terms: the quadratic case, Canadian
Appl. Math. Quarterly \textbf{13}, 89--106 (2005).

\bibitem{garnier1} R. Garnier, Sur des syst\`emes diff\'erentiels du
second ordre dont l'int\'egrale g\'en\'erale est uniforme, C. R. Acad.
Sc. Paris \textbf{249}, 1982--1986 (1960).

\bibitem{garnier2} R. Garnier, Sur des syst\`emes diff\'{e}rentiels du
second ordre dont l'int\'{e}grale g\'{e}n\'{e}rale est uniforme, Ann. \'{E}%
cole Norm. \textbf{77} (2), 123--144 (1960).

\bibitem{CCL} F. Calogero, R. Conte, and F. Leyvraz, New algebraically 
solvable systems of two autonomous first-order ordinary differential 
equations with purely quadratic right-hand sides (to be published)

\bibitem{schwarz1} Tobin A. Driscoll and Lloyd N. Trefethen, {\em
Schwarz--Christoffel mapping}. Vol. {\bf8} Cambridge University Press, 2002.

\bibitem{schwarz2} Thomas J. Higgins, An Epitomization of the Basic Theory
of the Generalized Schwarz--Christoffel Transformations as Used in Applied 
Physics, J. App. Phys. {\bf22}, 365--366 (1951).

\bibitem{schwarz}H.A. Schwarz, \"Uber einige Abbildungsaufgaben,
in {\em Gesammelte Mathematische Abhandlungen}, Berlin {\bf2}, 65--82 
(1890).

\bibitem{conway} John B. Conway, {\em Functions of one complex variable I} 
Vol.~159. Springer Science \&{} Business Media, 2012.

\bibitem{GS} P. Grinevich and P.M. Santini, Newtonian dynamics in the
plane corresponding to straight and cyclic motions on the hyperelliptic
curve $\mu ^{2}=\nu ^{n}-1$, $n\in {\mathbb{Z}}$: Ergodicity, isochrony and
fractals, Physica D \textbf{232}, 22--32 (2007).

\bibitem{category} John C. Oxtoby, {\em Measure and Category} (1971)
Springer-Verlag New York Heidelberg Berlin.

\bibitem{numtrian} Th. W. Ruijgrok, Periodic orbits in triangular billiards, Acta Phys. Polon. B {\bf22}
955--967 (1991)

\bibitem{trian4} H. Masur, Closed trajectories for quadratic differentials 
with an application to billiards, Duke Mathematical Journal, {\bf53} (2) 307--314 (1986).

\bibitem{trian1}  E. Gutkin, Billiards in polygons: survey of recent results, 
J. Stat. Phys.  {\bf83} (1--2), 7--26 (1996); E. Gutkin, Billiards in polygons, 
Physica {\bf19D} 311--333 (1986).

\bibitem{trian3} S. Kerckhoff, H. Masur, and J. Smillie, Ergodicity of billiard 
flows and quadratic differentials, Annals of Mathematics, {\bf124} (2) 293--311 (1986).

\bibitem{trian5} John Smillie, The dynamics of billiard flows in rational polygons of dynamical 
systems, {\em Dynamical systems, ergodic theory and applications}, {\bf100}, 360--382 (2000).

\bibitem{schwartz1} Richard E. Schwartz, Obtuse Triangular Billiards I: Near the (2, 3, 6)
Triangle, Experimental Mathematics, {\bf15} (2), 161--182 (2006).

\bibitem{schwartz2}  Richard E. Schwartz, Obtuse Triangular Billiards II: One Hundred
Degrees Worth of Periodic Trajectories, Experimental Mathematics, {\bf18} (2), 137--171 (2009).
\end{thebibliography}

\end{document}